\documentclass[aps,prl,twocolumn,groupedaddress,dvips]{revtex4}
\usepackage[dvips]{graphicx}

\begin{document}
%\include{zba1.bib}
%\documentclass[aps,prl,preprint,groupedaddress]{revtex4}
%\documentclass[aps,prl,preprint,superscriptaddress]{revtex4}
\title{Electron Transport in Metallic Grains}

\author{D. Davidovic, A. Anaya, A. L. Korotkov, M. Bowman}
\affiliation{Georgia Institute of Technology, Atlanta, GA 30332}
\email[]{dragomir.davidovic@physics.gatech.edu}
\author{M. Tinkham}
\affiliation{Harvard University, Cambridge MA 02138}
%\thanks{}


\date{\today}

\begin{abstract}
We discuss electron transport in individual nanometer-scale
metallic grains at dilution refrigerator temperatures.
In the weak coupling regime, the grains exhibit Coulomb blockade
and discrete energy levels.
Electron-electron interactions 
lead to clustering and broadening 
of quasiparticle states. Magnetic field dependences of tunneling resonances 
directly reveal Kramers degeneracy and Lande g-factors. In grains of Au, 
which have strong spin-orbit interaction, g-factors are strongly suppressed 
from the free electron value.
We have recently studied grains in the strong coupling regime. 
Coulomb blockade persists in this regime.
It leads to a suppression in sample conductance at zero bias voltage at low temperatures.
The conductance fluctuates with the 
applied magnetic field near zero bias voltage. We present evidence that
the fluctuations are induced by electron spin. This paper 
reviews the evolving progress in interpreting these observations. 
\end{abstract}

%\pacs{73.23.-b,73.63.-b,73.21.-b}

\maketitle

\section{Introduction}
In this paper, we review the results of electron transport measurements 
in metallic grains in weak electrical contact with leads. In these grains, 
the conduction electrons 
remain localized within the grain much longer than the time 
between bounces from the grain boundary. The time $\tau$ that 
a conduction electron remains localized within the grain is related to
the uncertainty in electron energy $\Gamma$ as
$\Gamma =h/\tau$. 
The Thouless energy of the grain encodes the notion of the time it takes an electron to explore the grain volume. It is given by $E_{Th}\sim\hbar v_F l/D^2$, where $v_F$, $l$, and $D$ are the Fermi velocity, elastic mean free path and the grain diameter.
Unlike the Thouless energy, $\Gamma$ is highly dependent on the contact resistance
between the grain and the leads.
The grain has the property that $\Gamma\ll E_{Th}$. This property differentiates 
the grain from higher-dimensional systems, where $\Gamma$ and $E_{Th}$ are
comparable. Because of this property, 
electron transport in grains is different from that in
higher dimensional systems. For example, in grains, the spin effect 
on magnetoconductance can be much stronger than the orbital effects. By comparison,
in higher dimensional systems, the Aharonov Bohm effect on magnetoconductance greatly exceeds the spin effect. Thus, normal metal grains have the ability to
spin-polarize and spin-analyze,~\cite{mandar} which opens a possibility to use grains as
devices in spintronics.

\begin{figure}
\includegraphics[width=0.45\textwidth]{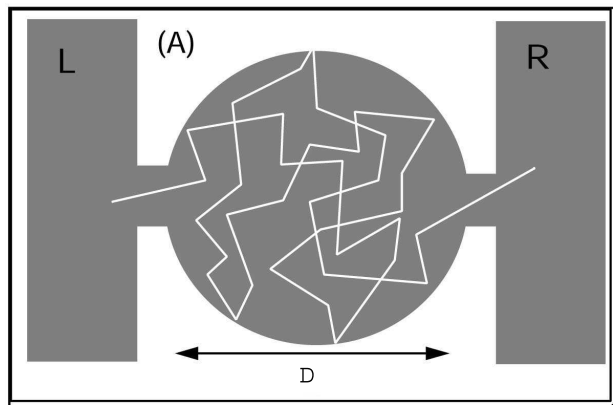}
\caption{A. Schematic of a metallic grain in weak electrical contact with the leads.
The geometry is chosen so that the conduction electrons bounce multiple times 
of the grain boundary during the course of their travel between the leads.\label{fig1}}
\end{figure}

\section{Closed and Open Grains}
Figure 1 shows a schematic of a sample containing one metallic grain 
connected between two 
electron reservoirs. Electron transport through the grain depends 
on the value of the contact 
resistances $R_{L}$ and $R_{R}$ between the grain and the leads. If both resistances are much larger than the resistance quantum, $R_Q=h/e^2$, the grain exhibits Coulomb blockade and discrete energy levels at low temperatures. In this regime, the grain is referred to here as a {\em closed} grain. 

Tunneling spectroscopy of energy levels in closed grains is difficult,
because
the grain diameter must be less than approximately $10nm$ in order to resolve the 
discrete energy levels at dilution refrigerator temperature. Such spectroscopic measurements were first carried out
by Ralph, et al. on grains of Al.~\cite{ralph} More recent measurements have extended this work
to grains of Au, Co, Cu, Ag and alloys of Al and Au. For a recent review see Ref.~\cite{delft}.

When $R=R_L+R_R$ is smaller than the
resistance quantum (roughly speaking),
$\Gamma$ becomes larger than the level spacing $\delta$. 
If $\Gamma >\delta$, the grain is referred to here as {\it open}.
In an open grain, both the energy spectrum and the
Coulomb blockade are washed out. 
The width in energy of
a charged state of the grain is $\tilde\Gamma \sim h/RC$, where $C=C_L+C_R$ is the sum of the junction capacitances. 
The ability to differentiate charged states of the grain is conditional on
$\tilde\Gamma < E_C =e^2/2C$. 

Two properties differentiate open grains from higher dimensional systems. First,
charging effects in open grains are not completely washed out.
Effectively, the charging energy
is exponentially reduced from the charging energy $E_C$ in closed grains, 
as $E_C^{eff}\sim E_C\exp (-\alpha R_Q/R)$, 
where $\alpha$ is a constant of order 1.~\cite{zaikin,nazarov}
$\alpha$ depends on the nature of the contacts. In tunneling contacts, $\alpha=0.5$~\cite{zaikin} and
in diffusive point contacts, the suppression is much stronger,  $\alpha=\pi^2/8$~\cite{nazarov}. 

Second, the correlation energy
of open grains is much smaller than the Thouless energy. The time that it takes
for an electron to traverse from one lead to the other is much longer 
than the time it takes to traverse through the grain volume. As described in the introduction, 
an electron bounces multiple times from the grain boundary in 
the course of its travel between the leads. Long localization time
is the reason that in a magnetic field, the spin-effect on transmission is stronger than the orbital effect. Spin up and spin down electrons begin to have different transmission when the Zeeman splitting becomes larger than the correlation energy. The directed area of an electron orbit
in the direction perpendicular to the magnetic field remains relatively small, despite the fact that  the localization time is long.~\cite{delft} As a result, the
Aharonov-Bohm flux remains weak even if the transmission has significant spin-dependence.

\section{Fabrication of Closed and Open Grains}

In standard metal deposition techniques, the grains typically nucleate at a certain center to center spacing, which depends on deposition rate and temperature. After the nucleation stage,
the deposition does not produce new grains. Instead, the grains grow in size. The 
grains tend to have a pancake shape with an irregular basis. 
Metal deposition is stopped before the grains form a percolating network.

The pioneering experiments by Ralph, et al.~\cite{ralph} were performed using a nanometer scale hole in an insulating Si$_3$N$_4$ membrane. 
The hole is used to select a single grain to establish a weak tunneling contact between two Aluminum
leads.
In the subsequent experiments by Davidovi\'c and Tinkham,~\cite{drago1} a careful shadow evaporation technique is used to create a nanometer scale
tunneling junction between two Aluminum leads. The grains are embedded inside the junction, and, 
electron transport is dominated by tunneling through a single grain. 

Recently, we have developed a new technique to create grains of 
circular shapes, in our laboratory at
the Georgia Institute of Technology. The grains are formed by
melting of a Au film on a $Si_3N_4$ substrate. 
A wide gold film of thickness 80nm is melted by applying a voltage pulse of
amplitude 10V and a
low source impedance. While melting,
the film breaks into electrically isolated droplets of Au, preventing
further current flow. Droplets are quenched by the substrate
which is at room temperature. Figure 2-A shows an irregular array of Au 
grains on the substrate, obtained by this melting process.

This principle is extended to quench a single gold droplet
between two larger gold electrodes. Figure 2 shows a gold grain
in weak electric contact with two Au leads. 
We verify through microscopy that no additional connections are formed outside the slit shown in Figure 2. Typical grain size is 20-40nm.
This device is obtained by careful melting of a point contact between two larger Au leads.
Melting is controlled by adjusting the amplitude and the source impedance of the voltage pulse.
Using surface Au migration techniques, the contact resistance 
between the grain and the leads can be tuned from $R\gg h/e^2$ 
to $R\ll h/e^2$. This ability permits us to explore grain physics in 
open-grain regime.

\begin{figure}
\includegraphics[width=0.45\textwidth]{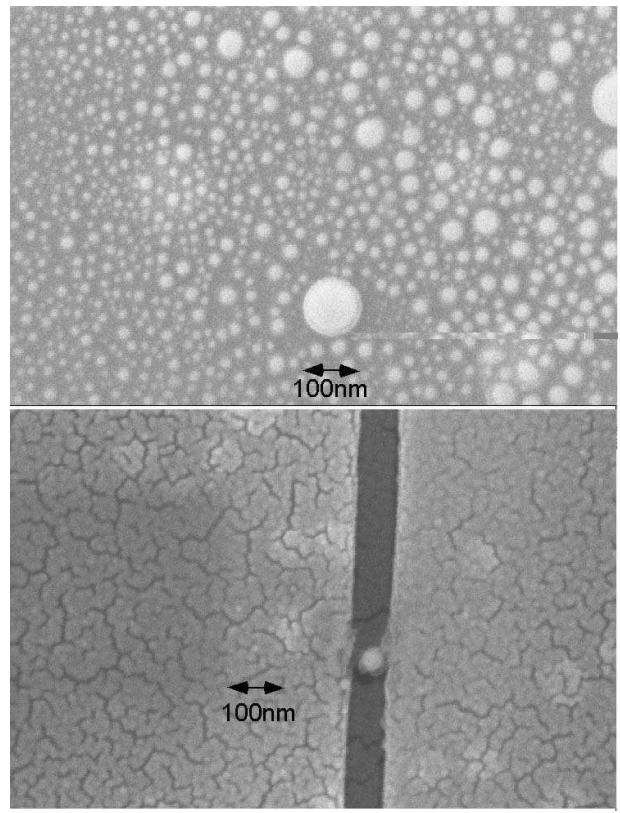}
\caption{A. 
An irregular array of circular Au grains formed by melting 
and quenching of Au film. B. 
One Au grain in weak electric contact 
with Au leads, formed by local melting and quenching.
\label{fig2}}
\end{figure}

\section{Spectroscopic Measurements of Discrete Energy Levels}

In metallic grains, the charging energy is typically much larger than the level spacing.
The reason is that the charging energy scales roughly as inverse of the grain area, 
and the level spacing scales as inverse of the grain volume. Since the surface to volume ratio is small, charging effects are observed even at temperatures at which the energy levels can not be resolved.

At temperatures where $\delta\ll k_BT\ll E_C$, the I-V curve of the grain displays 
Coulomb blockade. In good samples, the I-V curve is well described by the 
Orthodox theory of single charge tunneling.~\cite{likharev} This model permits evaluation of sample parameters, such as junction capacitances and resistances. Alternative techniques of parameter evaluations have also been developed - for details see original publications in Ref.~\cite{ralph,drago1}. 

As the temperature is lowered so that $kT\ll\delta$, individual energy levels of the grain 
become resolved. Figure 3 shows the I-V curve of a gold nanoparticle of diameter $\sim$ 5nm, at 30mK temperature. The steps in current correspond to discrete energy levels of the particle. Levels 1, 2, and 3 are indicated with arrows.

\begin{figure}
\includegraphics[width=0.5\textwidth]{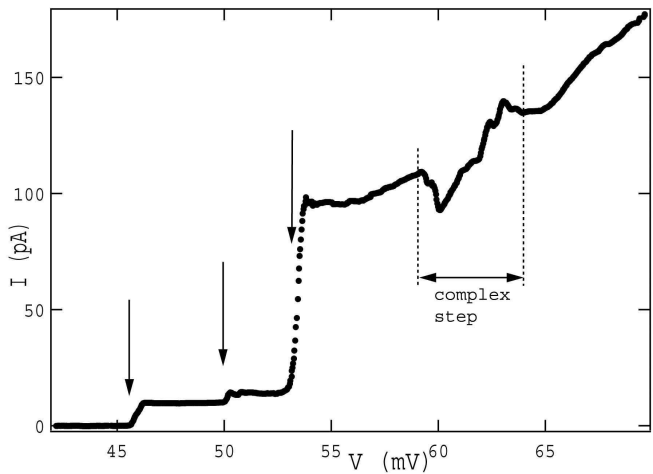}
\caption{Discrete electronic energy level spectrum in a ~5nm diameter Au nanoparticle. At low bias voltage, the current steps are simple. At larger bias voltage, the tunneling resonances (steps) become more complicated. A quasiparticle resonance at large voltage bias is mixed among several 
subresonances. Charging energy = $30meV$. Expected level spacing from particle size = $5meV$.\label{fig3}}
\end{figure}

\subsection{Effects of Electron-Electron Interactions}

A crude picture in which the eigenenergy of a many-electron system is
approximated by the occupation of a specific set of quasiparticle states, with simply 
additive energies is insufficient to explain the observed energy spectra. The first step beyond this picture was made by Agam et al.~\cite{agam} They pointed out the need to take into account of cross-terms in the Fermi-liquid energy expansion. In the Fermi liquid, the energy of 
a quasiparticle depends on what other quasiparticles are present. Thus,
instead of a single voltage associated with tunneling into a quasiparticle state, there will be a cluster of possible voltages, depending on what other quasiparticles are present. This possibility would not arise if one considered only equilibrium states at $T=0$, since only the lowest-energy configuration of excitations would be present. However, Agam et al.~\cite{agam} pointed out that, if successive tunneling events took place more quickly than relaxation from previous tunneling events took place, there would be a certain probability of nonequilibrium occupation numbers and hence of several possible eigenenergies for a given quasiparticle excitation. An important consequence of this model in its simplest form is that the lowest tunneling resonance should remain single, because a second tunneling event into a given level could not
take place until the level had been vacated, and when the lowest excited level is vacated, the grain is in its unique ground state.

The approach of Agam et al. was essentially perturbative, taking into account cross-terms in the Fermi liquid expansion. Somewhat earlier, Sivan et al.~\cite{sivan} had applied a
perturbative approach to estimate the level 
{\it width} which results from the lifetime limitation due to interelectronic scattering. Their conclusion was that level widths should increase with excitation energy,
and become as large as the level spacing when $E\approx E_{Th}$. This result implies that
above $E_{Th}$, the excitations are broad enough to blend into a continuum. Careful analysis of experimental measurements by Sivan et al. on diffusive semiconductor quantum dots supported this conclusion. 

The next major step in developing the theory of quasiparticle lifetimes was a nonperturbative treatment by Altshuler et al.~\cite{altshuler} Their approach was to map the problem of lifetimes into a problem of localization in the Fock-space of wavefunctions, analogous to the problem of Anderson localization on a Cayley tree. According to their analysis, the quasiparticle spectrum of a quantum dot separates into  four regimes with increasing excitation energy. 
The quasiparticle states are predicted to be sharp and single at low energy.
As the energy is increased, the resonances are sharp and clustered, then broad but resolvable, and finally forming an unresolved continuum above the Thouless energy. 

The early data by Ralph et al. were in quantitative agreement with the basic predictions 
that the tunneling resonances cluster and broaden with increasing energy. More recent 
measurements by Davidovi\'c and Tinkham have explored the progression from resolved narrow resonances into
effectively uniform tunneling density of states. Fig. 3 displays the current voltage 
characteristic of a 5nm Au grain. At low bias voltage, the tunneling steps are relatively sharp and simple. At 65 mV, a more complicated current threshold is found. It indicates
that at large bias voltage the tunneling resonances do not reflect single quasiparticles. 
The measurements by Davidovi\'c and Tinkham show that at voltages larger than the Thouless energy, quasiparticle states can not be resolved, in agreement with the earlier studies by Sivan et al.~\cite{sivan}. For further details, see Ref.~\cite{drago1}.

\subsection{Electron Spin}

The magnetic field dependence of eigenenergies in normal Al nanoparticles shows simple Zeeman splitting of two-fold degenerate energy levels.~\cite{ralph} The g-factor of spin doublets in Al has been measured to be $2\pm0.05$. The observation of two-fold spin degeneracy is a direct demonstration that the electronic states in normal metal grains are essentially the same as the quantum states of the non-interacting electron-in-a-box model, at least near the ground state.  The levels are filled sequentially, despite the fact that the grain contain several thousand strongly interacting conduction electrons. 

The applied magnetic field also affects the orbital energy of a discrete level.
In metallic grains, the orbital effect is usually much weaker 
than the Zeeman splitting.~\cite{delft} 
By comparison, in semiconducting quantum dots in perpendicular fields, the orbital effect greatly exceeds the spin effect. 

In Al nanoparticles, the effects of spin-orbit scattering can be neglected, because Al is a light element. In metallic grains of heavier elements, spin-orbit interaction reduces the g-factors of Kramers doublets. Spin orbit interaction induces mixing between pure spin up states and pure spin down states. 

\begin{figure}
\includegraphics[width=0.5\textwidth]{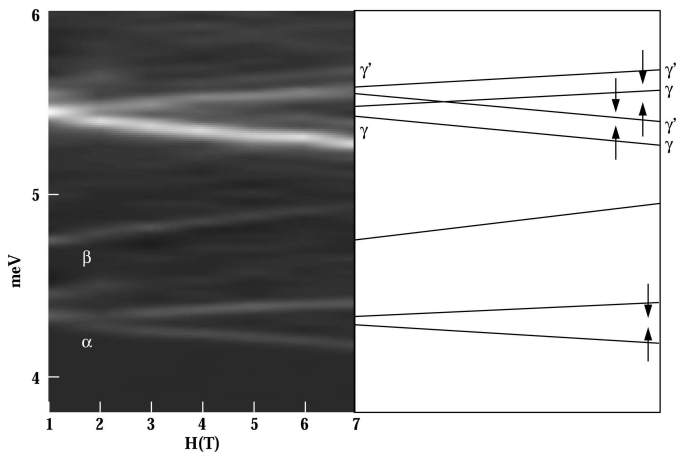}
\caption{Zeeman splitting of Kramers doublets in a  Au grain of diameter 9nm.}
\end{figure}

Salinas et al.~\cite{salinas} examined the role of spin orbit interaction induced by gold impurities in aluminum grains. They found that g-factors are reduced from 2. Salinas et al.
demonstrated avoided level crossing between Kramers doublets. 
Measurements in pure Au grains, made by Davidovi\'c and Tinkham, demonstrated
Zeeman splitting with g-factors much smaller than two. They found that in Au grains,
g-values range from $\sim 0.3$ to $\sim 0.5$. Figure 4 displays Zeeman splitting in a Au grain of diameter 9nm. 

Theories explaining g-factors
in small metallic grains have been developed by two groups at approximately the same time, by Matveev et al.~\cite{matveev} and Brouwer et al.~\cite{brouwer} 
Matveev et al. show that
the g-value reduction becomes significant when the spin-orbit scattering rate $\tau_{SO}^{-1}$ is comparable with the level spacing.
They predict that the g-factor  is distributed by the Maxwell
distribution among different Kramers doublets in the limit when $\tau_{SO}\delta/\hbar\ll 1$. The average g-value is 
\[
<g^2> =6 \delta\tau_{SO}/\pi\hbar+a l/D,
\]
where $a$ is a dimensionless constant determined by the geometry of the nanoparticle. 
The two terms represent the spin and the orbital contribution to
the g-factor. This equation shows that if the spin orbit scattering is strong, that is, $\delta\tau_{SO}/\hbar\ll 1$, then the g-factor is determined by the orbital contribution.
This contribution is of order 1 in a ballistic nanoparticle.

Hence, measurements of very small g-factors in Au grains are consistent with the theory 
if, 1), 
the spin-orbit scattering rate is much larger than $\delta/\hbar$, and 2), that the grains are
diffusive, that is, $l\ll D$. Short mean free path in these grains is surprising.
If a thick gold film is grown in identical conditions, the mean free path 
is much longer than the grain diameter studied by Davidovi\'c and Tinkham. 
For example, a g-factor of $0.3$ in a 9nm grain implies that the mean free path 
is less than $8\AA$, which is certainly not the case in bulk films.

It is possible that impurities such as water adsorb more easily into the grain than into bulk film. The grains are formed by nucleation, and they grow in size by capturing 
nearby Au atoms, which freely diffuse over the substrate surface. The substrate surface is
heavily contaminated by water molecules. This increases the probabilty that an impurity molecule 
is adsorbed inside the grain, possibly explaining the short mean free path. 

Brouwer at al. predict that the splitting of an energy level in a grain depends on the direction 
of the applied magnetic field, as a result of mesoscopic fluctuations.~\cite{brouwer} The anisotropy is described by the eigenvalues $g_j^2$ (j=1,2,3) of a tensor, corresponding to the g-factors along three principal exes. The anisotropy is enhanced by eigenvalues repulsion between $g_j$.

More recently, Petta and Ralph determined the effects of spin-orbit scattering on discrete energy levels in Copper, Silver, and Gold nanoparticles.~\cite{petta1} They determined the level to level fluctuations in the effective g-factor for Zeeman splitting. The statistics are found to be well described by the theoretical predictions. The strength of the spin-orbit scattering increases with atomic number and also varies between nanoparticles made of the same metal. 

Petta and Ralph have also measured the angular dependence, as a function of the direction of magnetic field,
for the Zeeman splitting of individual energy levels in copper grains.~\cite{petta2} They confirm the theoretical prediction by Brouwer at al., that the g-factors are highly anisotropic, with angular variations as large as five. Both the principal axes directions and g-factor magnitudes vary between different energy levels within one grain.

\section{Open Grains}

We have recently begun measuring open gold grains, using a new device geometry developed
in our laboratory at the
Georgia Institute of Technology.
For the remainder of this review, we present some of the most striking
effects we have discovered in these grains. 

In section 3, we showed that a single gold grain can
be captured between two gold leads. One device is shown in Fig. 2. 
These devices are very different from metallic quantum dots studied previously. 
First, the grain is formed by melting. It tends to have a spherical shape,
as opposed to the pancake shaped grains studied previously. 
We speculate that the mean free path in these grains is
significantly longer that the mean free path in grains formed by nucleation. 

Second, the contacts between the grain and the leads are not tunneling junctions.
In tunneling junctions, there are normally many channels, and every channel has a weak
transmission. In our new devices, the electrical contacts have few conducting channels, and every channel has a relatively large transmission, e. g. of order $10\%$. 
The contacts between the grain 
are sensitive to the motion of single gold atoms near the interface between the grain and the leads. This suggest that the electrical contact is closer to a {\it diffusive} metallic point contact than to a tunneling junction.

\begin{figure}
\includegraphics[width=0.5\textwidth]{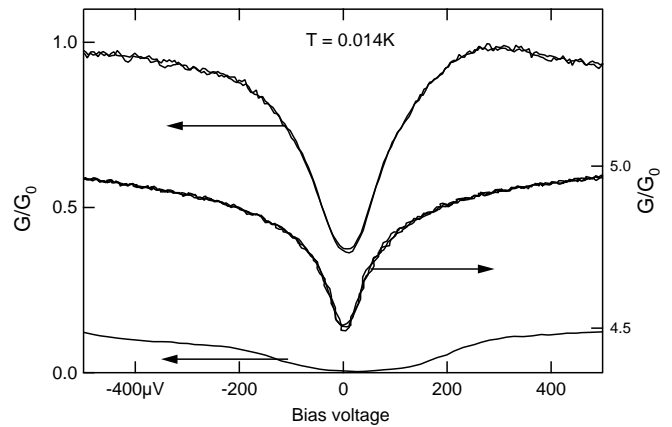}
\caption{Suppression in conductance of open grains at low temperatures. Three curves correspond 
to three different samples.
The zero bias conductance dip is a remnant of Coulomb blockade. $G_0=2e^2/h$}
\end{figure}

We study how the I-V curve at dilution refrigerator temperatures 
evolves as a function of total grain resistance. 
We find that if the room temperature resistance is larger than approximately $20k\Omega$,
samples 
display sharp Coulomb blockade at $T=0.015K$ with charging energy of order several meV. In approximately $30\%$
of those samples, the I-V curve is consistent with the Orthodox theory
of sequential electron tunneling through a single grain.~\cite{likharev}

The existence of Coulomb blockade demonstrates that 
the grain is weakly coupled to the leads.
If the sample resistance at room temperature is smaller than
approximately $10k\Omega$, Coulomb blockade is washed out at $T=0.014K$. Figure 5 shows 
differential conductance versus bias voltage
in three open grains. Typical grain diameter is 40nm. Typical charging energy (in weak coupling regime) is 4meV.
We show below that the weak zero-bias conductance dip
is a remnant of the Coulomb blockade. 

Open grains are characterized by comparing the I-V curves at low temperature
with the theory of
strong electron tunneling through mesoscopic metallic grains.~\cite{golubev} Describing the  details 
of sample characterization is beyond the scope of this summary.

\section{Coulomb Blockade in Open Grains}

The analysis of strong electron tunneling through metallic grains requires advanced theoretical methods.~\cite{flensberg,zaikin,golubev,falci,wang,konig,hofstetter} The ground state energy of the grain retains its periodic dependence on 
charge. Quantum fluctuations renormalize the grain parameters. If the grain is connected by tunneling junctions, 
the effective charging energy is suppressed by a factor of $exp(-G/2G_Q)$, where $G_Q=e^2/h$ and $G$ is the sum of the two contact conductances. Experiments in metallic islands connected by tunneling junctions
demonstrated the existence of charging effects 
for the values of $G$ exceeding $G_Q$.~\cite{devoret,chouvaev}

Charging effects are possible even if the grain is connected by metallic contacts, as long as they provide sufficient isolation of the grain from the leads. The total linear conductance between the grain and the leads at energy $E$ is $G=(e^2/h)\Sigma_n T_n(E)$, where $T_n(E)$ are the transmissions of the channels. $G$ fluctuates with energy, with a characteristic correlation energy $\sim h/\tau$. The effective charging energy of the grain takes the form~\cite{flensberg,nazarov}
\[
E_C^{eff}=E_C\Pi_n \sqrt{1-T_n}
\]

In diffusive contacts, averaging over distribution of $T_n$ over different channels
leads to~\cite{nazarov} 
\[
E_C^{eff}\sim E_C\exp(-\pi^2 <G>/8G_Q).
\]
$<G>$ is the total conductance between the grain and the leads, averaged over different
channels.

The quantity $<G>$ involves an additional averaging of the transmission coefficient
over a strip of energies of width $E_C$ around the Fermi level.~\cite{thomas,kamenev}.
This quantity almost does not fluctuate, since $E_C\gg h/\tau$.

This summary does not do justice to the subtlety of the theory. 
It is sufficient to show that the absence of any field dependence of the zero-bias conductance dip proves that the conductance dip in Fig. 5 is due to Coulomb blockade. It is not due to zero-bias anomaly in Altshuler-Aronov's sense, which would be split.~\cite{aronov}
We show in the next section that, after averaging over different impurity configurations, 
for any value of magnetic field, the differential conductance has a minimum at $V=0$.

\begin{figure}
\includegraphics[width=0.3\textwidth]{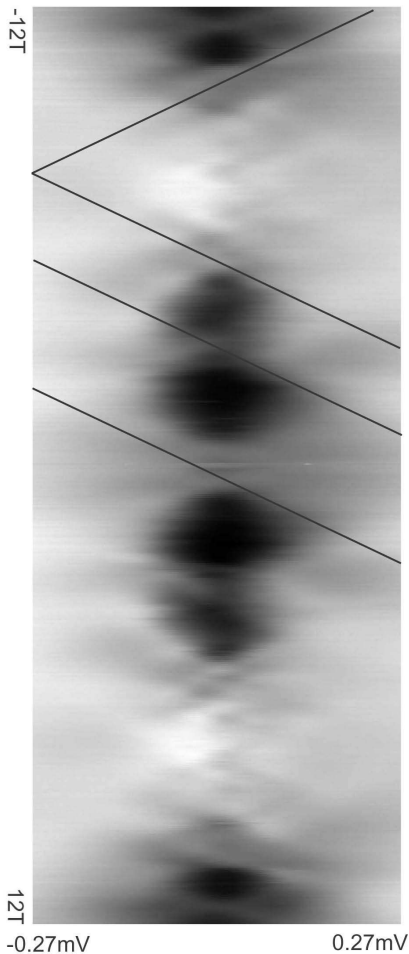}
\caption{Conductance of an open grain versus magnetic field and bias voltage. Darker=smaller conductance. Minimum conductance$\approx 0.22e^2/h$, maximum conductance $\approx 2e^2/h$.
Temperature is 0.015K.}
\end{figure}

\section{Conductance Fluctuations in Open Grains}

We show below that the conductance fluctuation in open grains
are based on electron spin.  
Essentially, in a magnetic field, spin up electrons and spin down electron have different wavelengths at the Fermi level.
In certain magnetic fields, spin up electrons may interfere constructively
when spin down electrons interfere destructively, or vice versa, resulting in net spin polarized current.

The influence of electron spin on electronic properties of open grains is demonstrated by tracing the fluctuations in conductance versus magnetic field and the bias voltage.
Fig 6 shows conductance fluctuations (CF) as a function of voltage and
magnetic field at $T=0.014K$ in one open grain with room temperature 
resistance of $\approx 14k\Omega$. 

The fluctuations of conductance 
versus magnetic field are strongly correlated with
fluctuations in conductance versus voltage. 
The image shows that there are
diamond shaped regions in the
parameter space
within which the conductance is suppressed.
In the figure, some of the diamond edges are highlighted with lines of the form
$eV=\pm 2\mu_B H+const$. 

The diamond edges are found to be consistent with the g-factors of bulk Au.~\cite{janossy} 
The detailed mechanism of how Zeeman splitting changes the I-V curve is not well understood yet.
In addition, in certain samples, the diamonds are absent; they are replaced with a 
dense network of lines of the form $eV=\pm 2\mu_B H+const$. Since the slope
of the lines is reproducible among samples, and since the slope
is effectively given by the g-factor of bulk $Au$ (g=2),
we are confident that the underlying interference effect is induced by the Zeeman splitting.

The fluctuations of conductance are highly sensitive to changes in the impurity configuration. Thermal
cycling leads to complete scrambling of the fluctuations at $T=0.015K$. The average conductance changes by less than $5\%$ with thermal cycling. By averaging conductance fluctuations over many different impurity configurations, we obtain a smooth background I-V curve, at any magnetic field. The zero-bias conductance dip has virtually no remaining magnetic field dependence after this averaging.

Theoretically, one must address the role of electron spin on 
the field dependence of the I-V curve, in the regime where the spin effects are much stronger than the orbital effects. Experimentally, we are adding a gate to our devices,
which will allow investigations of the fluctuations at zero bias voltage
without possible complications arising from nonequilibrium effects at finite bias voltage.

\begin{acknowledgments}
We are indebted to Michael Pustilnik, Leonid Glazman, Piet Brouwer, Yuval Oreg and Bertrand Halperin for
numerous useful discussions.
Part of this work was performed in part at the Cornell Nanofabrication Facility, (a member of the National
Nanofabrication Users Network), which is supported by the NSF, under grant ECS-9731293,
Cornell University and Industrial affiliates. We thank the microscopy facility at Georgia
Institute of Technology for the access to the scanning electron microscope.
This research is supported by the David and Lucile Packard Foundation 
grant 2000-13874 and the NSF grant DMR-0102960.

\end{acknowledgments}

\bibliography{zba1}

\end{document}